\documentclass[journal]{IEEEtran}
\usepackage{amsmath,amsfonts}
\usepackage{xcolor}
\usepackage{algorithmic}
\usepackage{algorithm}
\usepackage{array}
\usepackage{textcomp}
\usepackage{stfloats}
\usepackage{url}
\usepackage{verbatim}
\usepackage{graphicx}
\usepackage{cite}
\usepackage[caption=false,font=footnotesize,labelfont=rm,textfont=rm]{subfig}
\begin{document}

\title{\huge Phase Shift Information Compression in IRS-aided Wireless Systems: Challenges and Opportunities}

\author{Xianhua Yu, Dong Li,~\IEEEmembership{Senior Member,~IEEE}\vspace{-10mm}
\thanks{Xianhua Yu and Dong Li are with the School of Computer Science and Engineering, Macau University of Science and Technology, Macau 999078, China (emails: xianhuacn@foxmail.com; dli@must.edu.mo). }
}



\maketitle
\pagestyle{empty}  
\thispagestyle{empty}

\begin{abstract}
Intelligent reflecting surfaces (IRS) have emerged as a promising technology for future 6G wireless networks, offering programmable control of the wireless environment by adjusting the phase shifts of reflecting elements. However, IRS performance relies on accurately configuring the phase shifts of reflecting elements, which introduces substantial phase shift information (PSI) delivery overhead, especially in large-scale or rapidly changing environments. This paper first introduces the architecture of IRS-assisted systems and highlights real-world use cases where PSI delivery becomes a critical bottleneck. It then reviews current PSI compression approaches, outlining their limitations in adaptability and scalability. To address these gaps, we propose a prompt-guided PSI compression framework that leverages task-aware prompts and meta-learning to achieve efficient and real-time PSI delivery under diverse conditions. Simulation results show improved reconstruction accuracy and robustness compared to the baseline method. Finally, we discuss open challenges and outline promising directions for future research.
\end{abstract}

% \begin{IEEEkeywords}
% Intelligent reflecting surface, phase shift information delivery compression, deep learning
% \end{IEEEkeywords}

\section{Introduction}
Intelligent reflecting surfaces (IRS) have emerged as a key enabler for next-generation wireless communication systems, particularly in the context of 6G networks \cite{DLi1, DLi2, EShi, ECStrinati, CXWang}. By dynamically adjusting the electromagnetic response of a large array of nearly passive elements, IRS enables programmable control over the wireless propagation environment. This capability can significantly improve signal quality, extend coverage, and enhance both spectral and energy efficiency, without requiring additional active radio frequency (RF) chains.

To realize these benefits in practice, the system must accurately control the phase shifts applied by each IRS element. This control relies on the delivery of optimized phase shift information (PSI), which can be formulated as a matrix specifying the phase values denoted as a sequence of binary bits after quantization for each reflecting element. {The PSI is computed at the base station (BS) and transmitted to the IRS controller through a control channel. }Accurate and timely PSI delivery is essential for enabling the IRS to steer reflected signals in a manner that constructively enhances direct transmissions, thereby improving system capacity and reliability. {In this work, PSI refers specifically to the IRS phase configuration and does not include the phase of impinging signals or any feedback from the IRS to the BS.}

{However, PSI delivery introduces a significant challenge due to the limited bandwidth of the control channel between the BS and the IRS controller. The size of the PSI increases proportionally with the number of reflecting elements, which may reach several hundreds or thousands. Additionally, in dynamic environments involving user mobility or fast channel fading, the PSI must be updated frequently, creating frequent overhead. Unlike conventional systems where channel state information (CSI) is fed back from active users to the BS, it is imperative to consider the PSI delivery instead of the CSI feedback primarily due to the limited computational capacity of the IRS.}

Several recent studies \cite{XYu2, XYu1, ZLi, HFeng} have proposed deep learning-based PSI compression methods. While these approaches have shown their potential, they are typically trained under fixed system assumptions, such as constant compression ratios, static channel conditions, or specific SNR levels. As a result, they often generalize poorly in practical deployments where system parameters vary across time, users, and environments.

\par To address these limitations, we propose a prompt-guided PSI compression framework tailored for IRS-aided wireless systems. The design introduces a learnable Prompt Bank, where each prompt serves as a soft controller that dynamically modulates the encoder’s behavior based on task-specific metadata, such as compression ratio, SNR, or channel conditions. When explicit metadata is unavailable, the framework supports signal-based prompt matching to select the most appropriate control prompt directly from the input features.

To further enhance adaptability in unseen or evolving scenarios, we incorporate a meta-learning strategy that enables prompt updates using only a small number of support samples. This adaptation does not require retraining the encoder or decoder, ensuring a lightweight and scalable solution. The resulting architecture supports robust and efficient PSI compression across diverse and dynamic deployment environments.

To validate the effectiveness of the proposed framework, we conduct comprehensive simulations under various compression ratios, SNR levels, and channel conditions. Compared with baseline deep learning models trained under fixed assumptions, our prompt-guided approach demonstrates superior reconstruction accuracy, adaptability across scenarios, and robustness in unseen environments. These results confirm the practical viability of our design for real-time, scalable PSI delivery in IRS-assisted 6G networks.

\section{{Background and Challenges of PSI Delivery in IRS-aided wireless systems}}
\begin{figure*}
\vspace{-5mm}
    \centering
    \includegraphics[scale=0.45]{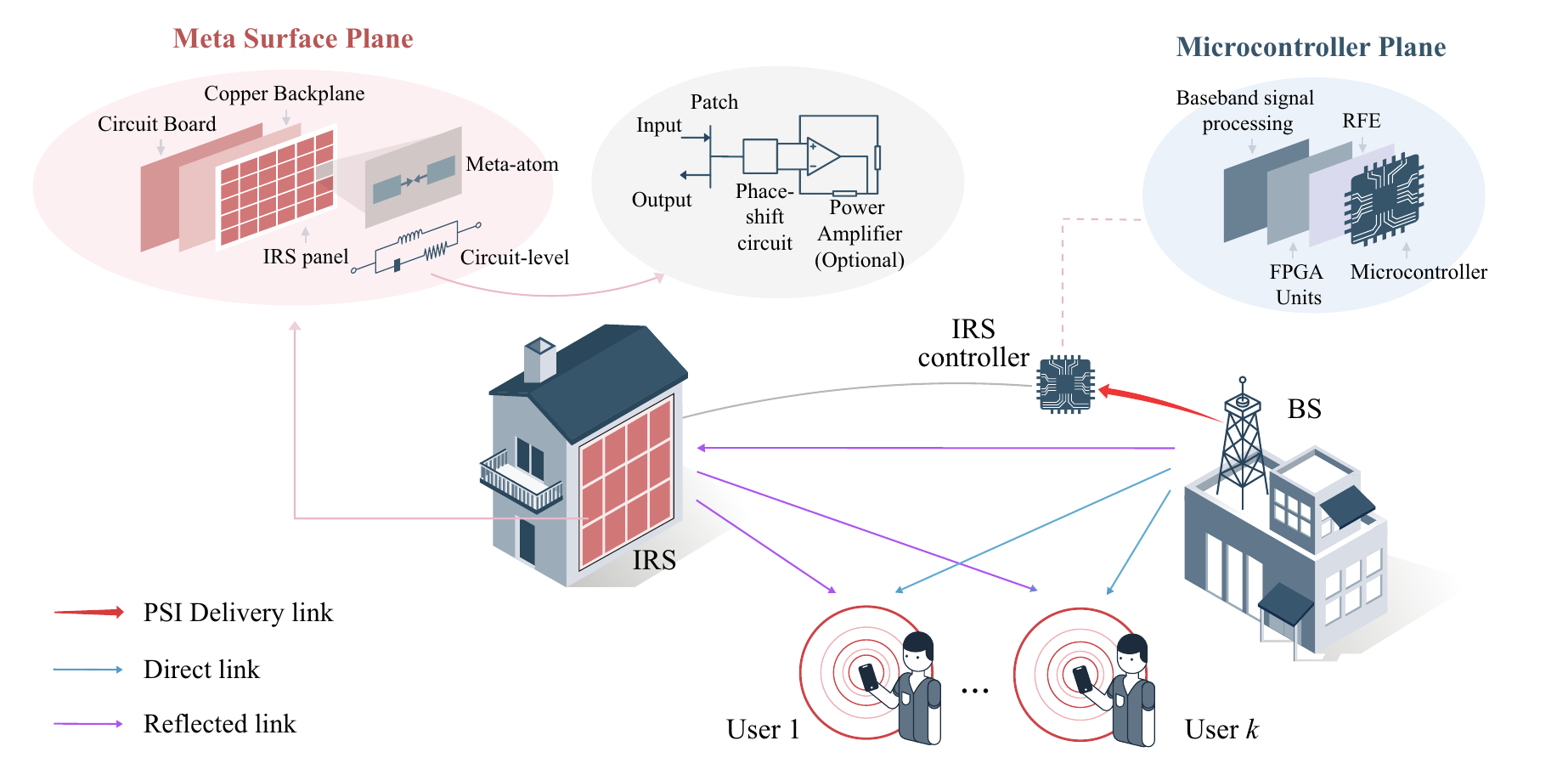}
    \caption{A typical IRS-aided wireless communication system and the structure of the IRS.}
    \label{fig1}
\end{figure*}
\par As illustrated in Fig.~\ref{fig1} the IRS typically comprises three distinct layers along with a control circuit. The outermost layer consists of metallic patches embedded on a dielectric substrate, which directly interact with incident signals. A copper backplate is placed behind this layer to prevent signal leakage. The innermost layer houses a control circuit board that adjusts the reflection amplitude and phase of each element, under the guidance of an embedded controller. In practice, this controller may be implemented using a field-programmable gate array (FPGA), and it communicates with the BS, access points (APs), or user devices through low-rate dedicated wireless links.

\par A typical IRS-aided wireless communication system, as shown in Fig.~\ref{fig1}, consists of a BS, multiple users each, and an IRS. The signal transmission process can be described as follows: for a given time interval, before data transmission, the BS needs to optimize the PSI. Then, the PSI is delivered to the IRS controller through a control link, and each element of the IRS adjusts the phase shift according to the received PSI. After the completion of the PSI delivery, signal transmission can begin.

\par The IRS offers a scalable and low-cost means to improve signal reliability and coverage. However, effective deployment hinges on the real-time, accurate delivery of PSI. This delivery process is challenged by several key factors as follows. First, as the number of IRS elements increases into the hundreds or thousands, the volume of PSI grows proportionally, significantly increasing signaling overhead.
Besides, each phase shift must be quantized before transmission. Higher quantization resolution further increases the amount of data.
Moreover, in dynamic or mobile environments, PSI must be frequently regenerated and delivered, resulting in repeated transmission overhead.

\par In certain application scenarios, PSI must be delivered multiple times in quick succession. For example, when a system alternates between the channel estimation stage and data transmission stage, each transition may require new PSI delivery. This creates substantial control signaling burdens, particularly in bandwidth-limited systems.

\begin{figure*}
\vspace{-5mm}
    \centering
    \includegraphics[width=0.9\linewidth]{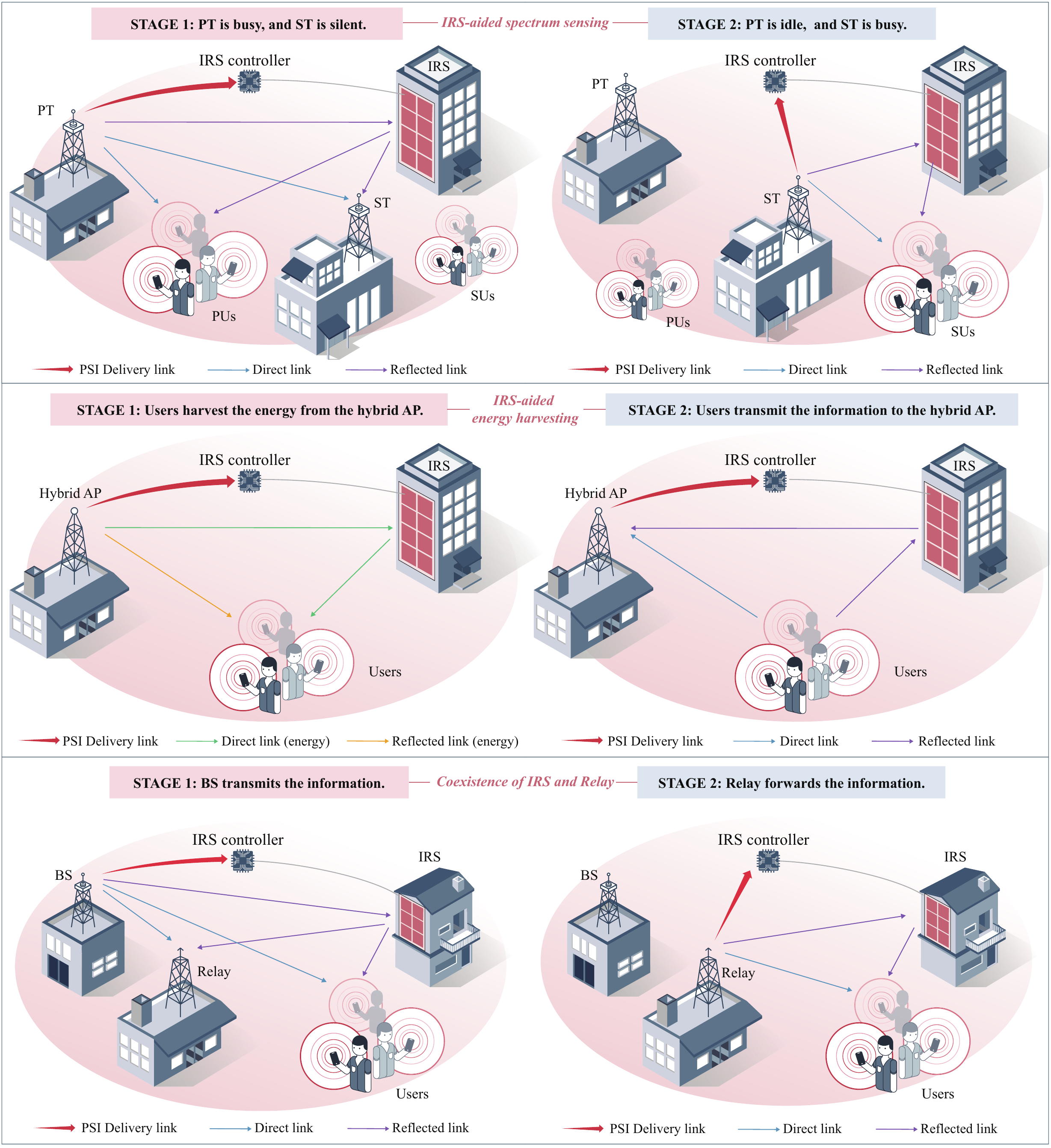}
\caption{The applications involve two stages of PSI delivery.}
\label{fig2}
\end{figure*}
% \begin{figure*}
% \vspace{-5mm}
% \centering
% % \subfloat[Channel estimation]{\label{fig2a}\includegraphics[scale=0.35]{1.eps}}\\
% % \hspace{-5mm}
% \subfloat[IRS-aided spectrum sensing]{\label{fig2b}\includegraphics[scale=0.358]{2.eps}}\\
% % \vspace{-5mm}
% \subfloat[IRS-aided energy harvesting]{\label{fig2c}\includegraphics[scale=0.358]{3.pdf}}\\
% \subfloat[Coexistence of IRS and Relay]{\label{fig2d}\includegraphics[scale=0.358]{4.eps}}
% \caption{The applications involve two stages of PSI delivery.}
% \label{fig2}
% \end{figure*}
 
\subsection{Spectrum sensing}

\par Spectrum sensing is used to enhance the spectrum efficiency (SE) of traditional bands to cope with the limited spectrum resources as IoT devices increase rapidly. By detecting the status of the primary network, unlicensed secondary users can fully utilize frequency bands when the primary network is idle. However, detection may fail when the SNR of the primary user (PU) signal at the secondary user (SU) receiver is low. The detection accuracy can be improved by the IRS, which strengthens the PU signal SNR.

\par As shown in Fig.~\ref{fig2}, the IRS-aided spectrum sensing system consists of a primary network and a secondary network and operates in two stages in different time slots: the spectrum sensing stage and the data transmission stage. In the spectrum sensing stage, the primary transmitter (PT) first delivers the PSI to the IRS controller, and the secondary transmitter (ST) continuously monitors the primary network and receives the signal from the PT through two links: the direct link (PT-ST) and the reflected link (PT-IRS-ST). In the data transmission stage, the ST starts to transmit information to secondary users when the primary network is detected to be idle. Before information transmission, the PSI needs to be delivered from the ST to the IRS controller.

\par Moreover, in spectrum sensing tasks that support ultra-reliable low-latency communication (URLLC) applications such as mission-critical control or vehicular networks, both sensing and transmission phases must occur within ultra-low latency budgets. Even minor delays or inaccuracies in PSI delivery between stages can disrupt timely access to the channel or degrade detection reliability. This highlights the need for ultra-efficient, low-latency PSI delivery mechanisms in such scenarios.

\subsection{Energy Harvesting}

\par In practice, given the high cost of conventional power-grid-based solutions, most small devices are battery-powered. However, limited battery power can potentially shorten network lifetimes and degrade the quality of service. This power shortage can be addressed by wireless power transfer (WPT), which scavenges energy from energy-carrying signals radiated by energy transmitters. However, WPT faces the challenge of efficiency degradation with increasing distance. With assistance from the IRS, the system can increase energy efficiency while still guaranteeing the quality of service (QoS) for each IoT device.
 
\par A typical IRS-aided wireless power communication network (WPCN) is shown in Fig.~\ref{fig2}. Before the stage of data transmission from users to the BS, the users need to be involved in the energy harvesting stage. Specifically, the users on the left side are in the energy harvesting stage. To maximize energy efficiency, the PSI should be delivered to the IRS controller from the hybrid AP. The users on the right side have already charged enough to meet the transmission power requirement and can transmit information to the hybrid AP. However, some devices, such as sensors, may transmit information frequently and require frequent WPT, leading to a high PSI delivery overhead where a large amount of PSI must be fed back. In such scenarios, to guarantee both energy efficiency and QoS, reducing the PSI delivery overhead is necessary.

\subsection{Cooperative Relaying}
\par The relay complements the signal processing capabilities of the IRS, enabling the network to facilitate sophisticated transmission strategies. Additionally, the IRS can enhance the network transmission quality in a cost-effective manner when combined with relays.

\par The DF relay with half-duplex mode coexists with the IRS, serving both the users and the BS in the IRS-aided relay system, as shown in Fig.~\ref{fig2}. This system consists of two transmission stages. In the first stage, the BS transmits the information to the user and the relay, which is aided by the IRS to reflect the signal, and the relay amplifies/decodes the information. In the second stage, the relay forwards the information to the user, aided by the IRS to improve the transmission quality.

\par The PSI delivery is essential in both stages, leading to high overhead and potentially hindering the benefits of the convergence of IRS and relay, making it necessary to reduce the overhead.
\begin{table*}[t]

\centering
\caption{Comparison of PSI Compression Solutions}
\begin{tabular}{|p{1.8cm}|p{4cm}|p{4cm}|p{4cm}|}
\hline
\textbf{Solution} & \textbf{Main Idea} & \textbf{Advantages} & \textbf{Limitations} \\
\hline
\textbf{PSCDN} & CNN autoencoder with denoising module & Simple implementation; noise-robust PSI recovery & Limited scalability in dynamic or complex environments  \\
\hline
\textbf{GAPSCN} & Global attention with deep skip-connected network & High reconstruction accuracy; attention-aware compression & High computational complexity; not suitable for IRS-side decoding \\
\hline
\textbf{S-GAPSCN} & Asymmetric design with lightweight decoder & Deployable on resource-limited IRS controllers; good trade-off & Slightly reduced accuracy compared to GAPSCN \\
\hline
\textbf{ACFNet} & Adaptive compression via learned policy network & Flexible compression ratio control; improved sum-rate adaptation & Unstable training due to non-differentiable mask selection \\
\hline
\textbf{PSFNet} & Knowledge-based shared index compression & High compression ratio; compact memory use & Relies on accurate index decoding; sensitive to channel mismatch \\
\hline
% \textbf{Tensor (PARAFAC)} & Low-rank factorization with Kronecker product & Low overhead; interpretable structure in LoS settings & Performance degradation in NLoS conditions & LoS-dominated IRS deployments \\
% \hline
% \textbf{Tensor (Tucker)} & Multi-linear Tucker decomposition & Strong in NLoS scenarios; flexible parameter tuning & Slightly higher delivery overhead; complex tuning required & NLoS-dominant or irregular PSI environments \\

\end{tabular}
\label{table:psi_comparison}
\end{table*}
\subsection{Channel estimation}
\par The performance of an IRS-aided wireless communication system is highly dependent on the accuracy of the channel state information (CSI) between the BS and the IRS, as the optimization of the PSI is based on the CSI. In the channel estimation stage, after the BS transmits the PSI to the IRS controller, it continually sends pilot signals to the BS, which are reflected by the IRS. The CSI of the cascaded channel between BS-IRS and IRS-user is estimated at the BS based on the received pilots. In the data transmission stage, the PSI is delivered from the BS to the IRS controller first. Then, the BS transmits information to the user aided by the IRS. Since each stage requires PSI delivery from the BS to the IRS controller and the two stages switch rapidly between each other, a substantial amount of PSI needs to be delivered in each time slot, leading to high PSI delivery overhead.

\par This becomes especially critical in high-frequency systems such as THz communications, where extremely short wavelengths and narrow beamwidths require frequent, high-precision PSI updates. Any delay or inaccuracy in PSI delivery can lead to beam misalignment and severe performance degradation. In such cases, fast and accurate PSI control becomes a limiting factor for system reliability and throughput.

\section{Conventional Solutions for Reducing PSI Overhead}
\par In this section, common solutions for reducing PSI overhead are detailed as follows, and a summary is shown as table \ref{table:psi_comparison}.
\subsubsection{PSCDN} 
\par It is worth noting that the challenge of PSI delivery overhead was investigated for the first time in \cite{XYu1}, where an autoencoder-based neural network named phase shift compression network (PSCDN) was proposed to address the issue. PSCDN is built on a convolutional neural network (CNN) with a ReLU activation function but without using a batch normalization (BN) layer to maintain consistent distribution. This is because the BN layer would nonlinearly transform the input to a Gaussian distribution, while the PSI follows a uniform distribution. Additionally, in PSCDN, a denoising module is introduced in the decoder to alleviate the noise effect. The denoising module consists of two parts: a neural network part and a subtraction part. The neural network estimates the noise, while the subtraction part performs a subtraction between the reconstructed PSI and the estimated noise. Through this noise estimation and subtraction process, the reconstructed PSI achieves high accuracy.

\subsubsection{GAPSCN} 
\par The global attention phase shift compression network (GAPSCN) was proposed in \cite{XYu2} to achieve reliable PSI compression performance. GAPSCN employs additional layers to enhance performance and uses skip connections to avoid the vanishing gradient problem. Moreover, to further improve performance, a novel global attention mechanism is proposed and utilized in GAPSCN. This global attention mechanism is designed to enhance performance by focusing on the important parts of the PSI. Unlike conventional attention modules, the global attention mechanism adopts both macroscopic and microscopic perspectives. Therefore, the global attention mechanism is capable of calculating attention maps across three dimensions—the channel, spatial, and joint channel-spatial dimensions—which emphasize a large number of meaningful features. Although GAPSCN achieves prominent performance with the help of a deeper neural network and global attention, the computational complexity should be considered a bottleneck in practice.

\subsubsection{S-GAPSCN} 
\par Considering the practical situation where the IRS side lacks computational resources, as the IRS controller is a simple FPGA that cannot handle high computational complexity tasks, the DL model needs to be redesigned to meet the requirements of limited computational resources. To this end, the simplified GAPSCN (S-GAPSCN) was proposed in \cite{XYu2}. This model adopts an asymmetric structure, where the architecture of the decoder is much simpler than that of the encoder, accommodating the computational resource constraints. To offset the performance degradation arising from the decoder's architecture simplification, a low-complexity module, JAAMSN, has been introduced in the S-GAPSCN decoder. Specifically, the JAAMSN module strives to emphasize meaningful features along the joint channel-spatial dimension to improve model performance. Additionally, by employing a multi-scale structure, the JAAMSN module is able to alleviate the effects of AWGN.

\subsubsection{ACFNet} 
\par To adaptively control the compression rate while maximizing the downlink effective achievable sum rate and ensuring its performance, an adaptive compression autoencoder-based model named ACFNet was proposed in \cite{ZLi}. The adaptive feedback compression is based on a policy network in the encoder, which aims to provide a mask vector, a one-hot vector to define the compression rate. Specifically, in the policy network, a Softmax layer is used to obtain the probabilities of the data. These probabilities are then transformed into a one-hot vector using a Gumbel-Softmax layer and a Lambda layer. An element-wise product is then performed between the quantized PSI and the mask vector. Through this element-wise product, adaptive PSI compression can be achieved. The main drawback of ACFNet is that the policy network cannot be trained using the backpropagation process, resulting in an unstable training process.

\subsubsection{PSFNet}
\par A knowledge-based autoencoder, PSFNet, was proposed in \cite{HFeng} to further reduce the compression rate. This model differs from others by incorporating a learnable knowledge base shared between the encoder and the decoder. Compared to conventional models, the compression process in PSFNet involves two stages: a conventional neural network compression stage and a novel shared knowledge-based compression stage. In the shared knowledge-based compression stage, the similarity between the feature vector of the compressed PSI and the knowledge-based vector is examined. The compression rate can be further reduced by only transmitting the index of the knowledge-based vector that is most similar to the feature vector of the compressed PSI. In the decoder, assuming the index is received correctly at the IRS side, the compressed PSI can be reconstructed based on the shared knowledge base. The decoder then reconstructs the original PSI based on the received compressed PSI. However, achieving further compression rates may be challenging due to the unrealistic assumption that the index will be transmitted losslessly through the wireless channel.

\section{Prompt-Guided PSI Compression Framework}
\begin{figure*}[t]
\vspace{-5mm}
    \centering
    \includegraphics[width=0.9\linewidth]{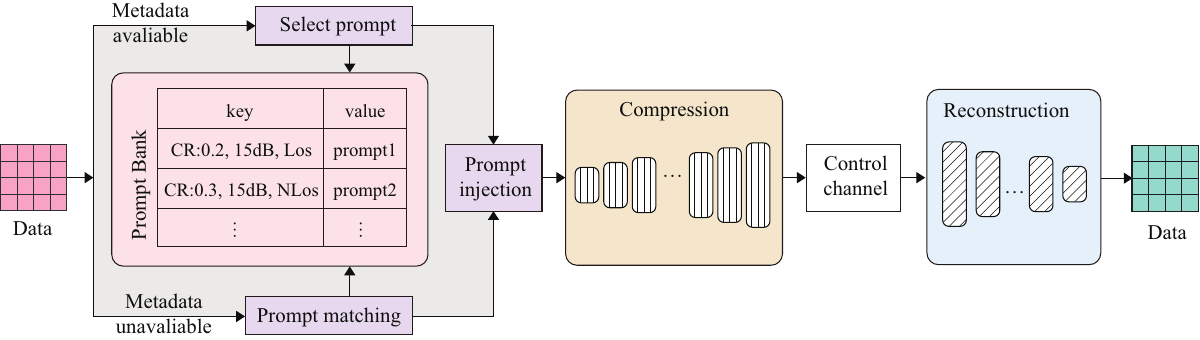}
    \caption{The proposed prompt-guided framework.}
    \label{fig3}
\end{figure*}
Most existing PSI compression methods are trained under fixed compression ratios (CRs) and predefined channel conditions. As a result, they struggle to generalize and often degrade in performance when exposed to dynamic or mismatched scenarios. To address these limitations, we propose a prompt-guided compression framework that dynamically adapts the encoder’s behavior based on current CR, channel condition, and signal characteristics, without retraining.

As illustrated in Fig.~\ref{fig3}, our framework consists of three main components: a Prompt Bank, a Prompt Matching Module, and an asymmetric autoencoder.

The prompt bank holds a set of learnable prompt vectors, each corresponding to a specific task scenario. These are stored as key–value pairs, where the keys represent environmental metadata such as CR, SNR, and channel type, and the values are the associated prompts. Rather than adjusting model weights, these prompts serve as soft controllers, conditioning the encoder’s internal behavior—such as attention and feature abstraction pathways—based on task context.

The prompt matching Module selects the appropriate prompt. When metadata is available, the system directly retrieves the corresponding prompt. In cases where such information is unavailable, the PSI input is embedded and compared against prompt keys using cosine similarity to find the best match based on signal content.

The autoencoder adopts an asymmetric architecture to suit practical deployment. A high-capacity Transformer-based encoder compresses PSI under dynamic task conditions, with behavior modulated by the selected prompt. The decoder, on the other hand, is lightweight and optimized for resource-constrained IRS controllers. Despite its simplicity, it effectively reconstructs PSI from task-aware compressed representations.

A key feature of our framework is its support for variable-rate compression. Prompts activate latent gating mechanisms or adaptive pooling, enabling the encoder to generate compressed outputs of different lengths with a single encoder–decoder pair. This eliminates the need for multiple task-specific models and simplifies deployment.

The step-by-step process of the framework is as follows:
\begin{enumerate}
    \item Input PSI reception: A PSI matrix is received as input.
    \item Prompt selection: If task metadata is available, the corresponding prompt is retrieved directly from the prompt bank using the key. If metadata is unavailable, the input PSI is passed through the Prompt Matching Module, which embeds the signal and compares it with stored prompt keys to select the most relevant prompt.
    \item Prompt injection: The selected prompt is injected into the encoder, influencing attention and feature abstraction behavior.
    \item Compression: The encoder compresses the PSI matrix into a task-aware latent representation. The prompt also controls the output dimensionality, using latent gating or adaptive pooling to match the target compression ratio.
    \item Reconstruction: The lightweight decoder reconstructs the PSI from the latent vector.
\end{enumerate}

To ensure adaptability to unseen scenarios, we incorporate a few-shot adaptation mechanism based on meta-learning. During training, prompts are optimized using multiple task episodes, while encoder and decoder weights remain fixed. At inference time, prompt vectors can be fine-tuned using only a small support set, enabling fast adaptation without full retraining.

In summary, our prompt-guided framework provides a unified, scalable, and flexible PSI compression solution. It dynamically adjusts to changing environments, supports multiple CRs and channel types, and reduces overhead without compromising performance. This makes it particularly suitable for real-world 6G applications such as THz beamforming and URLLC, where fast, accurate, and adaptive PSI delivery is essential.

\section{Simulation Results}
\begin{figure}
\vspace{-5mm}
\centering
\subfloat[NMSE across CRs under NLoS and 15 dB SNR]{\label{fig4a}\includegraphics[scale=0.5]{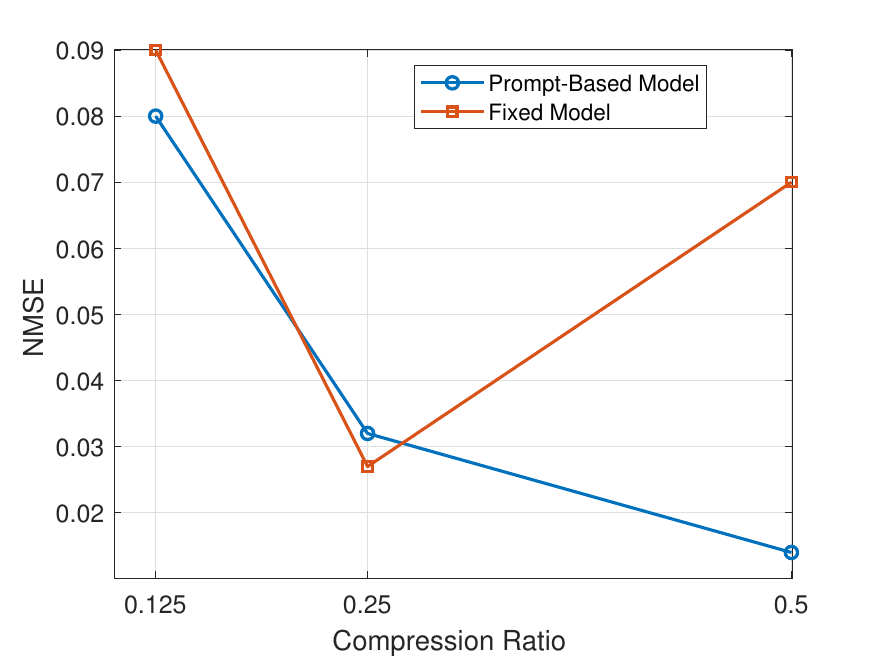}}\\
% \vspace{-5mm}
\subfloat[Generalization performance across LoS/NLoS channels and varying SNRs at CR = 0.25.]{\label{fig4b}\includegraphics[scale=0.5]{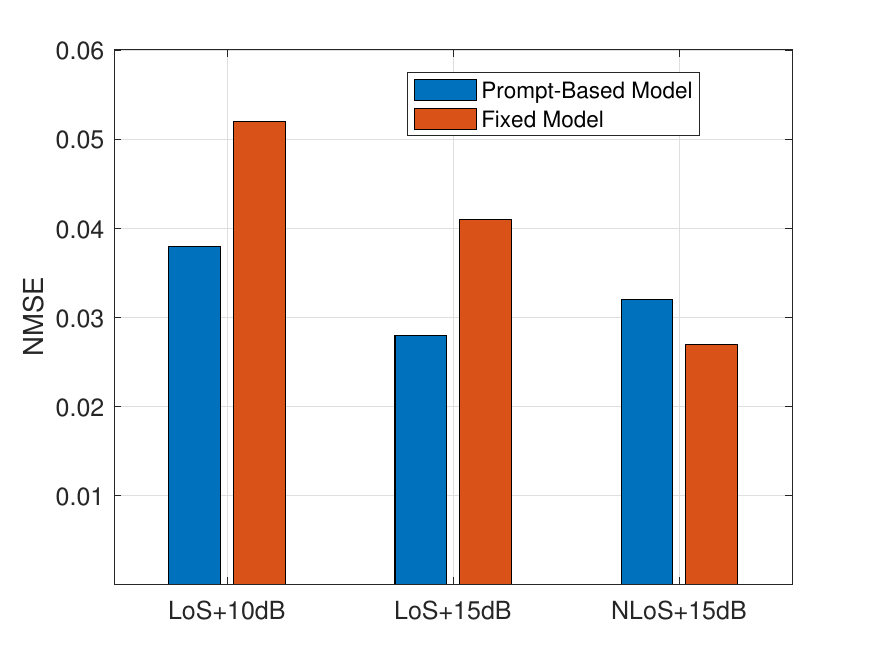}}\\
\caption{Comparison of NMSE performance for Different Methods. The baseline model trained under a single configuration, specifically, a CR of 0.25, an NLoS channel, and an SNR of 15 dB.}
\label{fig4}

\end{figure}
To evaluate the effectiveness of our proposed framework, we conduct comprehensive simulations across varying compression ratios, SNR levels, and channel conditions. We compare our method against a baseline autoencoder trained under a single configuration, specifically, a CR of 0.25, an NLoS channel, and an SNR of 15 dB.

As shown in Fig.~4(a), we evaluate performance under CRs of 0.125, 0.25, and 0.5. Our prompt-based model achieves robust reconstruction accuracy across all compression levels. In contrast, the baseline model performs well only at its training condition (CR = 0.25) but fails to generalize to other compression settings. This highlights the adaptability of our framework, which maintains low NMSE with a single encoder–decoder pair.

In Fig.~4(b), we fix CR = 0.25 and test across different SNRs (10 dB and 15 dB) and channel types (LoS and NLoS). Our framework maintains strong performance across all settings, while the baseline shows significant degradation, particularly under LoS conditions or unseen SNR levels. This confirms the generalization strength of prompt-based conditioning.

Unlike the baseline, which requires separate models for each task setting, our prompt-guided design handles all scenarios with one unified model. The prompt mechanism enables contextual awareness without retraining, and the lightweight decoder ensures compatibility with real-world IRS hardware.

These results collectively validate our approach: the proposed framework delivers robust, flexible, and efficient PSI compression, making it well-suited for dynamic 6G systems that demand real-time responsiveness and resource-awareness.

\section{Open Issues and Future Directions}
While this paper introduces a novel prompt-guide PSI compression framework, further exploration is needed to meet the growing demands of future IRS-aided wireless systems. In particular, intelligent control under dynamic conditions, resource constraints, and security threats introduces several important research challenges. Below, we outline key open directions based on three promising methodologies: continual learning\cite{LWang}, semantic communication\cite{XLuo}, and latency-aware design\cite{HXie}.

\subsection{Adaptation under dynamic conditions}
IRS-assisted systems deployed in real-world environments must operate under non-stationary and evolving conditions, such as user mobility, dynamic channel states, and varying deployment scenarios. In these settings, continual learning offers the promise of enabling PSI compression models to incrementally adapt over time, improving robustness without requiring full retraining.

However, applying continual learning in this context raises two critical challenges. First, the problem of catastrophic forgetting becomes pronounced: neural models may overwrite previously learned compression strategies when adapting to new channel conditions, especially in scenarios with seasonal or repetitive variations. Second, many existing continual learning methods depend on memory replay mechanisms or auxiliary modules, which are difficult to deploy on the resource-constrained IRS side. These components often assume a level of computational capability and storage that passive or semi-passive IRS hardware cannot provide. Furthermore, storing past PSI or environment data introduces privacy risks, particularly when such data reflects sensitive user behavior or spatial information.

To make continual learning practical for PSI compression, future research should focus on developing lightweight, buffer-free learning algorithms that can operate within hardware limitations. Additionally, methods based on regularization or parameter-isolation may offer efficient alternatives to data replay. Privacy-preserving learning frameworks that avoid raw data storage or transmission while still supporting robust adaptation are also essential for trustworthy deployment in real-world systems.

% \subsection{Limited training data}
% \par Generally, training a DL model often requires a large amount of training and validation data to ensure reliable performance. Additionally, as the complexity of tasks increases, the amount of required data may also increase. However, in real-world scenarios, collecting and storing training PSI data can be challenging, leading to insufficient data. In such situations, how can we achieve a well-trained DL model with satisfactory performance?

% \par To this end, two approaches from few-shot learning, meta-learning \cite{THospedales} and generative models \cite{SBond-Taylor}, can be used to fill the gap. Meta-learning, also known as learning to learn, involves training a model to understand the meta-knowledge of a task. Thus, the meta-learning model is capable of adapting well to new tasks and environments that were never encountered, even with limited training data. In contrast, generative models can synthesize similar PSI data to enlarge the training dataset. By doing so, enough training data is generated to meet the requirements for model training.

\subsection{Semantic compression for PSI}
While traditional PSI compression techniques emphasize bit-level accuracy, semantic communication introduces a new paradigm that focuses on transmitting the underlying meaning of control information. By selectively discarding redundant or context-irrelevant elements, semantic communication can significantly reduce overhead while preserving the functional intent of PSI. In this context, applying semantic techniques to PSI delivery could yield superior efficiency and bandwidth savings, especially in dense or high-frequency deployments.

Nevertheless, several limitations must be addressed before semantic communication can be practically applied to IRS systems. One of the primary challenges is computational complexity. Semantic encoders and decoders typically rely on deep neural architectures that demand high processing power and memory, making them incompatible with the lightweight, low-power design of IRS controllers. Another concern is vulnerability to adversarial perturbations. Semantic models are known to be sensitive to even minor changes in input, which can drastically affect the decoded meaning. In the context of IRS control, such deviations could lead to incorrect phase shift settings, system degradation, or even malicious misbehavior.

Advancing semantic communication for PSI will require the development of compressed and distilled semantic models that can operate in real-time under constrained hardware. Equally important are robust training methodologies that can withstand input noise, environmental uncertainty, and potential attacks. Research should also explore task-aware semantic encoding, wherein PSI is abstracted and prioritized based on its operational relevance to system performance, rather than being uniformly compressed.

\subsection{Timely PSI delivery in latency-critical systems}
In emerging applications such as URLLC and THz beamforming, timely PSI updates are essential. Delays in PSI delivery can result in beam misalignment, service outages, or suboptimal channel utilization. Consequently, ensuring latency-aware PSI control is vital for supporting these high-performance scenarios.

Despite this need, most current PSI compression methods are designed with a focus on reconstruction accuracy alone, without explicit consideration of latency. However, in real-time environments, the PSI delivery process must balance accuracy with timing and resource usage. In other words, encoding strategies must be optimized not just for precision, but also for speed and responsiveness under varying network and channel conditions.

Future research in this direction should aim to develop delay-sensitive PSI compression mechanisms that adaptively prioritize critical phase elements based on their impact on performance and timing constraints. Adaptive PSI scheduling algorithms, capable of dynamically selecting which information to transmit under given latency budgets, will also be crucial. Furthermore, cross-layer optimization strategies that coordinate PSI delivery with medium access and network-layer policies can help achieve end-to-end delay guarantees in practical deployments.

\section{Conclusion}
In this article, we examined the critical issue of high PSI delivery overhead in IRS-aided wireless systems. We first reviewed representative deep learning-based PSI compression methods and analyzed their limitations in dynamic and resource-constrained environments. To address these challenges, we proposed a novel prompt-guided PSI compression framework that supports adaptive, scalable, and real-time control across diverse conditions. Our simulation results demonstrated the framework’s effectiveness in improving compression flexibility and robustness. Finally, we outlined several open issues and future research directions, including continual learning, semantic compression, and latency-aware design. We hope this work will inspire further research into scalable and intelligent PSI control for next-generation IRS-assisted networks.
% \section*{Acknowledgments}
% This should be a simple paragraph before the References to thank those individuals and institutions who have supported your work on this article.

% \newpage

 

% \end{IEEEbiography}



\vfill

\end{document}